\documentclass[letterpaper]{jpconf}
\usepackage{epsf,graphicx,amsmath,iopams}
\usepackage[square,sort&compress]{natbib}

\def\sun{\odot}

\def\kmps{$\mathrm{km\,s^{-1}}$}
\def\rsun{$R_{\sun}$}
\def\cone{{\it Case 1}}
\def\ctwo{{\it Case 2}}
\begin{document}
\title{Modeling the Near-Surface Shear Layer: Diffusion Schemes Studied With CSS}
\vspace{-0.2truein}
\author{Kyle Augustson$^1$, Mark Rast$^2$, Regner Trampedach$^1$, Juri Toomre$^1$}
\address{$^1$ JILA, $^2$ LASP, and Dept. of Astrophysical \& Planetary Sciences, University of Colorado, Boulder, CO, USA, 80309}
\ead{kyle.augustson@colorado.edu}

\begin{abstract}
As we approach solar convection simulations that seek to model the interaction of small-scale 
granulation and supergranulation and even larger scales of convection within the near-surface 
shear layer (NSSL), the treatment of the boundary conditions and minimization of sub-grid scale 
diffusive processes become increasingly crucial. We here assess changes in the dynamics and the 
energy flux balance of the flows established in rotating spherical shell segments that capture 
much of the NSSL with the Curved Spherical Segment (CSS) code using two different diffusion 
schemes. The CSS code is a new massively parallel modeling tool capable of simulating 3-D 
compressible MHD convection with a realistic solar stratification in rotating spherical shell 
segments.
\end{abstract}

\vspace{-0.25truein}
\section{Introduction\label{sect1}}
The solar differential rotation profile exhibits prominent radial shear layers near the top 
and bottom of the convection zone \citep{thomp03}. The near-surface shear layer (NSSL) occupies 
the upper 5\% of the Sun by radius, whereas the tachocline begins near the base of the convection 
zone. The dynamics of the NSSL are governed largely by vigorous granular-scale convection that is 
driven by radiative cooling and large thermodynamic gradients. The collective interaction of 
these granular-scale flows (average sizes of 1~Mm, lifetimes of 0.2~hr) is a major component in 
the formation of supergranular (15-35~Mm, 24~hr) and mesogranular (5-10~Mm, 5~hr) scales
\citep{rast03,nord09}. 

Given this wide range of spatial and temporal scales, we currently cannot simultaneously 
model hundreds to thousands of supergranules, solar granulation, and deeper flows. However, 
we may still be able to characterize the influence of granulation and deep global flows on 
the NSSL by coupling CSS with global convection models from below \citep{miesch08} and with 
surface convection simulations from above \citep[e.g.][]{rempel09,nord09}. Such a coupling, 
whether it is statistical or direct, requires a careful treatment of diffusion and boundary 
conditions \citep{august10}. For instance, low diffusion is necessary to preserve the spatial 
structure and advective timescales of the downdrafts flowing into the CSS domain from the 
surface convection above. Given that our grid is five times coarser than that of typical surface 
convection simulations, this is no easy task.

We have conducted two numerical simulations in a $20^\circ$ square patch centered on the equator 
that encompass most of the NSSL and some of the deep interior, rotating at the solar rate. These 
simulations explore the effects of turbulent-eddy and slope-limited diffusion schemes with an open 
lower radial boundary and closed upper boundary; as such they are identical except for the diffusion 
scheme. The governing equations and numerical approach used in solving them are briefly discussed in 
\S \ref{sect2}. The turbulent-eddy and slope-limited diffusion schemes are detailed in \S \ref{sect3}. 
The dynamics of the flows established in these simulations are examined in \S \ref{sect4}.

\section{Formulating the Problem\label{sect2}}
The spherical segment domains used in our simulations involve large portions of the Sun's inherent 
spherical geometry, which is necessary to properly capture the effects of rotation on supergranular 
scales. We use the Curved Shell Segment (CSS) code to model the 3-D dynamics below the solar 
photosphere. CSS is a modeling tool that solves the compressible Navier-Stokes equations in rotating 
spherical segments \citep{august10}. To simulate the larger scales of motion that are likely to occur 
in the solar convection zone, a large-eddy simulation (LES) model is employed. The scales that are 
not explicitly computed in these simulations are parametrized and included in a sub-grid scale model 
of turbulent transport and diffusion. The governing equations solved in CSS are:

\begin{figure}[t]
\begin{center}
\includegraphics[width=\textwidth]{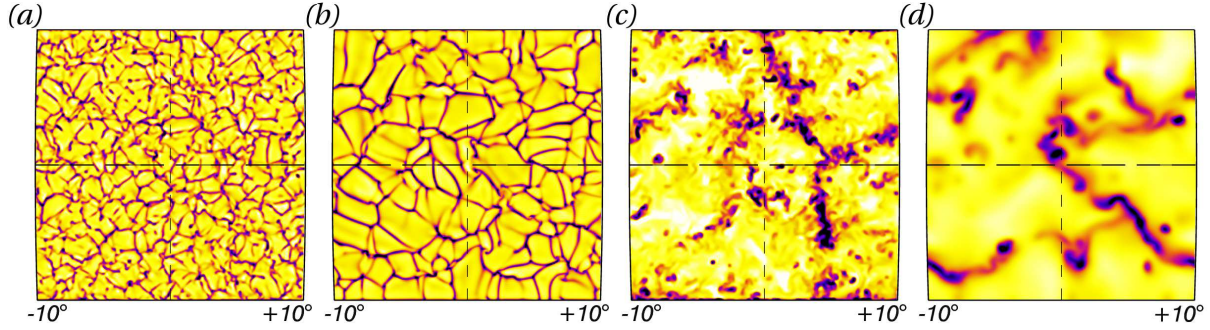}
\vspace{-0.25truein}
\caption{Radial velocities ($u_r$) at 0.99~\rsun\ in (a) the slope-limited diffusion \ctwo, and (b) the 
  turbulent-eddy diffusion \cone, and at 0.95~\rsun\ in (c) \ctwo\ and (d) \cone. Dark tones denote 
  downflows and light tones denote upflows. Scaling values are -1.34~\kmps, 0.59~\kmps\ for (a) and (b) 
  and -0.73~\kmps, 0.24~\kmps\ for (c) and (d), where positive $u_r$ points outward.}
\end{center}
\vspace{-0.25truein}
\end{figure}

\vspace{-0.4truein}
\begin{center}
   \begin{eqnarray}
     \frac{\partial \rho}{\partial t} + \nabla \cdot \left( \rho \mathbf{u} \right) & = & 0, \\
     \rho \left[ \frac{\partial \mathbf{u}}{\partial t} + \left(\mathbf{u} \cdot \nabla \right) \mathbf{u} \right] & = & -\nabla P - 
     \rho g \hat{\mathbf{r}} + \nabla \cdot \bar{\mathcal{D}} + 2 \rho \mathbf{u}\times \mathbf{\Omega}
          + \rho \Omega^2 \mathbf{R}, \\
     \rho T \left[ \frac{\partial S}{\partial t} + \left(\mathbf{u} \cdot \nabla \right) S \right] & = & 
          \nabla \cdot \left[\kappa_S \mathbf{\nabla} S' + \kappa_0 \frac{\partial \bar{S}}{\partial r} \hat{\mathbf{r}} 
	    + \kappa_r \mathbf{\nabla} T \right] + \Phi, \\
     \bar{\mathcal{D}}_{ij} & = & \mu \left[ \bar{e}_{ij} - \frac{1}{3} \nabla \cdot \mathbf{u} \bar{\delta}_{ij} \right].
   \end{eqnarray}
\end{center}

\vspace{-0.1truein}
The symbols $\rho$, $\mathbf{u}$, $P$, $T$, $S$, $\bar{\mathcal{D}}$, $\bar{e}$, $\bar{\delta}$ are 
the density, velocity, pressure, temperature, specific entropy, viscous stress tensor, stress tensor, 
and identity tensor respectively; $\Phi$ is the viscous heating term; the entropy perturbations are 
$S'=S-\langle S \rangle$; the equation of state is $P = \rho^{\gamma} e^{S/C_V}$; $\mathbf{\Omega}$ is the 
angular velocity of the rotating frame; $\mu$ and $\kappa_S$ are the turbulent-eddy diffusion of 
momentum and entropy, and $\kappa_r$ is the radiative thermal diffusion; $g$ is the local acceleration 
due to gravity, where $\mu$, $\kappa_S$, $\kappa_r$, $\kappa_0$, and $g$ are functions of radius only. 

The governing equations (Equations 1-3) are evolved on a uniform spatial mesh. Temporal discretization 
is accomplished using an explicit fourth-order accurate Runge-Kutta time-stepping scheme. The spatial 
derivatives are computed using a modified sixth-order compact finite difference scheme. A 3-D domain 
decomposition divides the full spatial mesh into sub-domains. The boundary information necessary to 
compute spatial derivatives is passed between nearest-neighbor sub-domains using MPI, while 
computations are shared among the master and slave cores within a supercomputer node using OpenMP.

Currently there are two diffusion schemes implemented in CSS, a turbulent-eddy diffusion (TED) 
(Equation 3-4) and the slope-limited diffusion (SLD) described in (\S \ref{sect3}). In the TED scheme, 
the momentum and entropy diffusivities ($\mu$ and $\kappa_S$) are calculated based upon the desired 
Rayleigh number at the upper boundary with the constraint that the Prandtl number and the dynamic 
viscosity be constant throughout the domain. There are no directly comparable diffusive parameters in 
the SLD scheme as the eddy diffusion terms have been dropped. One may still control the level 
of diffusion, however, with specific choices of which slope-limiter is employed and what characteristic 
velocity is used at cell interfaces. Figure 1 depicts snapshots of typical flows at two radii from the 
two diffusion schemes.

\section{Slope-Limited Diffusion In CSS\label{sect3}}
Slope-limited diffusion has many possible formulations, and we have chosen to use a scheme similar to 
that found in \citep{rempel09}. SLD is based upon a piecewise linear reconstruction within a finite 
volume (cell) centered on each gridpoint of the solution at each time step. The linear reconstruction 
leads to a solution that has discontinuities at the cell edges. SLD essentially acts to minimize these 
discontinuities so that the numerical scheme remains stable. While this model of diffusion is not as 
physically motivated as the eddy diffusion model, it holds many computational advantages. Indeed, with 
SLD in CSS, a higher level of turbulence is achieved for a given resolution and there is no need to 
evaluate second-order derivatives, which decreases memory usage by a factor of three and the 
execution time per time step by a factor of two.

The slope of the linear reconstruction of the solution is given by a ratio of the downwind and upwind 
cell-center differences ($r_i = \Delta_l u_i / \Delta_r u_i$). This slope is ``limited'' by a 
function $\phi(r_i)$ that belongs to a class of slope limiters that yield total variation diminishing 
solutions \citep{leveq02}. As in \citep{rempel09}, we use a linear combination of two such slope limiters, 
the minmod and superbee limiters. Using the reconstructed slope, values of the primitive variables are 
computed at the cell edges as $u_i^e = u_i\pm\frac{1}{2}\phi\left(r_i\right)\Delta_r u_i$. The diffusive 
flux at a cell edge is $f_i^e = \frac{1}{2} c_i^e g_i^e \beta_i^e \delta u_i^e$, where $g_i^e$ is a 
geometric factor and $c_i^e$ is a characteristic velocity at a cell edge, and $\delta u_i^e$ is the 
difference of the left and right reconstructed values at a cell-edge. To avoid artificial steepening, 
a local diffusion coefficient $\beta_i^e$ is constructed such that when $\delta u_i^e \Delta_e u_i > 0$, 
$\beta_i^e = (\delta u_i^e/\Delta_e u_i)^2$ and is zero otherwise. This preserves steep gradients and 
yields a fourth-order upwind diffusion scheme \citep{rempel09}.

With the diffusive fluxes at each edge of a given cell as above, the diffusion at the center of a cell 
is $D_i = \left(f_i^r-f_i^l\right) / \Delta x_i$, where $\Delta x_i$ is the grid spacing. This is added 
to the solution after the full Runge-Kutta time step as $u_i = u_i + \Delta t D_i$. Within our simulations, 
using the true sound speed in the $c_i^e$ proves to be overly diffusive. So, for the computation of the 
$c_i^e$, the sound speed throughout the computational domain is fixed to a fraction ($\sim 1/10$) of the 
surface sound speed.

\section{Comparing Two Diffusion Schemes \label{sect4}}
The simulations encompassing a $20^{\circ} \times 20^{\circ}$ domain are relatively low resolution with 128 radial 
mesh points and 256 points in latitude and longitude. A stellar evolution code is used to establish a realistic 
initial stratification for the simulations. Since a perfect gas is assumed, the He and H ionization zones cannot 
currently be simulated. Thus, the impenetrable upper boundary is taken to be 0.995~\rsun\ in order to exclude 
most of the radial extent of these zones. The permeable lower boundary is placed at 0.915~\rsun, yielding a density 
contrast across the domain of 400. Each case has the lowest diffusivities that allow numerical stability in the 
TED (\cone) and SLD (\ctwo) regimes. Using a constant $\nu$ with depth in the TED case is rather restrictive in that 
the diffusion is larger than necessary away from the upper boundary, but for this preliminary analysis it is 
sufficient to characterize the differences between the two schemes. 

\begin{figure}[t]
\begin{center}
\includegraphics[width=\textwidth]{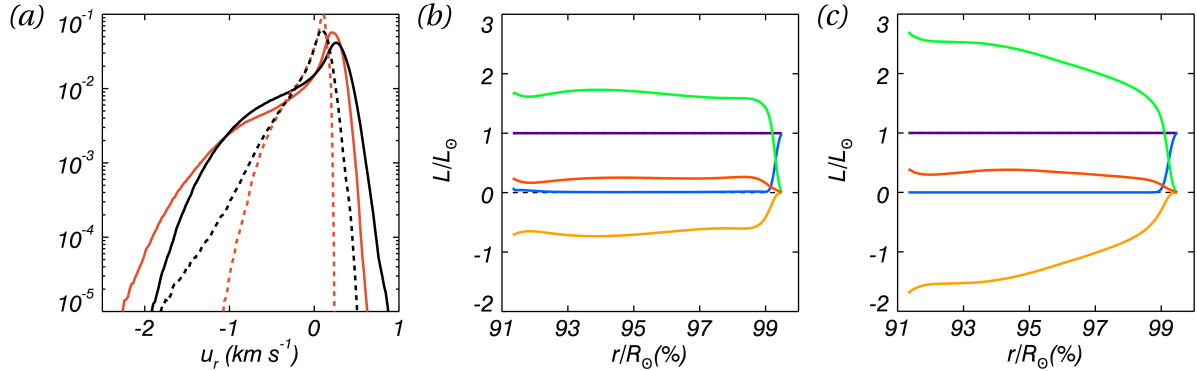}
\vspace{-0.25truein}
\caption{(a) PDFs of radial velocities at 0.99~\rsun\ (solid line) and 0.95~\rsun\ (dashed), for \ctwo\ (black) 
  and \cone\ (red). Time-averaged radial energy flux balances for \cone\ (b) and \ctwo\ (c), with total flux in 
  purple, radiative flux (blue), enthalpy flux (green), kinetic energy flux (orange), and acoustic flux (red).}
\vspace{-0.25truein}
\end{center}
\end{figure}

Turning to the radial velocity patterns sampled in Figure 1, there is a striking shift toward smaller spatial 
scales in \ctwo\ with considerably less power at large scales. At 0.99~\rsun\ the peak of the spatial power 
spectrum in \cone\ is around 40~Mm. In \ctwo, however, there is a broad peak at scales around 20~Mm, more 
closely matching the supergranular scales of the Sun. As one descends deeper into the simulations (Figures 1c,d), 
the flows appear significantly more turbulent in \ctwo\ than in \cone. 

Indeed, the downflowing structures of \ctwo\ become much more narrow at depth with many more isolated plumes. 
The greater turbulence also manifests itself in larger extrema of each of the physical variables. For example, 
the radial velocities at 0.95~\rsun\ in \ctwo\ have extrema that are twice that of \cone\ (Figure 2a). In \cone, 
a precipitous drop in the upflowing radial velocity probability density distribution (PDF) outside of about 0.1~\kmps\ 
is indicative of just how uniform the upflows are. In stark contrast, the radial velocity PDF of \ctwo\ has twice 
the dynamic range, emphasizing that the upflows are indeed much more structured than those of \cone. While the radial 
mass flux PDF has higher velocity wings that contribute only one percent to the overall distribution, the bimodal 
nature of the temperature perturbation PDF has the effect of significantly amplifying the importance of these outlying 
values. Thus, the net effect of the larger wings in these PDFs is to shift the mean of the enthalpy flux in
\ctwo\ to a higher value. The additional velocity in both the horizontal and radial flows greatly alter the kinetic 
energy distribution, as in \ctwo\ where the mean kinetic energy has nearly doubled and correspondingly increased
the magnitude of the kinetic energy flux (Figure 2c).

A tentative next step is to have both diffusion schemes active at the same time. This will allow a systematic 
study on the convergence of solutions as the TED coefficients are lowered to solar values and the SLD takes over 
as the primary diffusion.

\ack
The authors thank Neal Hurlburt and Marc DeRosa for most helpful discussions and their contributions to code 
development and implementation. This work has been supported by NASA grants NNX08AI57G and NNX10AM74H. The 
computations were carried out on Pleiades at NASA Ames with SMD grant g26133.

\bibliography{soho2010_augustson}

\end{document}